\documentclass[final,5p,times,twocolumn]{elsarticle}
\usepackage{lineno,hyperref,amsmath,amsthm,amssymb,amsfonts,ragged2e,color,subfig}
\journal{``Heliyon"}
\begin{document}
\begin{frontmatter}
\title{Modulational instability of dust-ion-acoustic mode and associated rogue waves in a non-extensive plasma medium}
\author{N. K. Tamanna$^{*,1}$, J. Akter$^{**,1}$, N. A. Chowdhury$^{\dag,2}$, and A. A. Mamun$^{\ddag,1,3}$}
\address{$^1$Department of Physics, Jahangirnagar University, Savar, Dhaka-1342, Bangladesh\\
$^2$Plasma Physics Division, Atomic Energy Centre, Dhaka-1000, Bangladesh\\
$^3$Wazed miah science research centre, Jahangirnagar University, Savar, Dhaka-1342, Bangladesh\\
e-mail: $^*$tamanna1995phy@gmail.com, $^{**}$akter277phy@gmail.com\\
$^{\dag}$nurealam1743phy@gmail.com, $^\ddag$mamun\_phys@juniv.edu}
\begin{abstract}
The modulational instability of dust-ion-acoustic (DIA) mode and associated
rogue waves in a three component dusty plasma system (containing
inertial warm ion and negatively charged dust fluids along with inertialess $q$-distributed
electrons) has been theoretically investigated. A  nonlinear Schr\"{o}dinger  equation (NLSE)
has been derived by employing the reductive
perturbation method. It is observed that the dusty plasma system under
consideration supports the fast and slow DIA modes, and that
the dispersion and nonlinear coefficients of the NLSE determine the parametric regimes not only for the modulationally
stable and unstable fast DIA mode, but also for the formation of the DIA rogue waves.
The parametric regimes for the modulational instability of the fast DIA mode,
and the criterion for the formation of the DIA rogue waves have been
found to be significantly modified by the effects of the relevant plasma
parameters, particularly, mass and charge state of ion and dust species, number
density of the plasma species, and non-extensive parameter $q$, etc.
It is found that the modulationally stable parametric regime decreases (increases) with
the increase in the value of positive (negative) $q$.  The numerical analysis has also
shown that the nonlinearity as well as the amplitude and width of the rogue waves increases (decreases)
with the mass of positive ion (negative dust grains) while decreases (increases)
with the charge state of the positive ion (negative dust). The applications of our present work
in  space (viz., Earth ionosphere, magnetosphere, molecular clouds, interstellar medium,
cometary tails, and  planetary rings, etc.) and laboratory plasmas have been pinpointed.
\end{abstract}
\end{frontmatter}
\section{Introduction}
\label{2sec:Introduction}
The presence of highly charged massive dust grains in a three component dusty plasma medium (DPM) along with electron and ion
in space plasmas (viz., Earth ionosphere \cite{El-Labany2018}, asteroid zones \cite{Eslami2013}, magnetosphere \cite{El-Labany2018},
protostellar disks \cite{Eslami2013}, molecular clouds \cite{Eslami2013}, interstellar medium \cite{El-Labany2018},
cometary tails \cite{Eslami2013}, and planetary rings \cite{El-Labany2018}, etc.) as well as laboratory
plasmas \cite{Bandyopadhyay2008,Heinrich2009,Popel2001} not only change the dynamics of the plasma medium but also
change the characteristic of the various kinds of nonlinear electrostatic disturbance,
viz., dust-acoustic (DA) waves (DAWs) \cite{Sahu2012a,Ferdousi2017,Sahu2012b}, DA solitary waves (DASWs) \cite{Sahu2012a}, DA shock
waves (DASHWs) \cite{Sahu2012a,Ferdousi2017}, DA double layers (DADLs) \cite{Sahu2012b}, dust-ion-acoustic (DIA)
waves (DIAWs) \cite{El-Labany2018,Eslami2013},  DIA solitary waves (DIASWs) \cite{El-Labany2018,Eslami2013},
DIA shock waves (DIASHWs), DIA double layers (DIADLs), DIA cnoidal waves (DIACWs), and  has also mesmerized
many plasma physicists to examine the propagation of nonlinear electrostatic perturbation.

A number of authors have considered non-extensive $q$-distribution in describing non-equilibrium plasma species
(due to the existence of long range interactions like gravitational and coulomb interactions) in the long tail space
environments such as asteroid zones \cite{C1,C2,C3,C4,C5,Tsallis1988,Akter2015,Bacha2012},
Earth ionosphere \cite{Tsallis1988,Akter2015,Bacha2012}, and interstellar medium \cite{Tsallis1988,Akter2015,Bacha2012}, etc.
In the $q$-distribution, the parameter $q$ describes the non-extensivity of the species, and the
species are to be considered super-extensive species for the limit $q<1$ while the species are to be considered  sub-extensive
species for the limit $q>1$ \cite{Tsallis1988,Akter2015,Bacha2012,Renyi1955}. El-Labany \textit{et al.} \cite{El-Labany2018}
examined DIA multi-solitons in three components DPM. Sahu and Tribeche \cite{Sahu2012a} investigated the nonlinear
properties of the DASWs and DASHWs in presence of the non-extensive plasma species. Akter \textit{et al.} \cite{Akter2015}
theoretically and numerically analyzed the nonlinear propagation of the ion-acoustic waves in a plasma having non-extensive plasma
species, and reported that the saturation of the curve for the low and high frequency is totally depend on the non-extensive parameter of the plasma species.
Bacha \textit{et al.} \cite{Bacha2012} considered inertial ions, inertialess electrons, and immobile dust grains to
study DIASWs in a three components DPM, and found that the profile becomes narrower with $q$ for
sub-extensive limit of $q$ while wider with $q$ for super-extensive limit of $q$.

The nonlinear Schr\"{o}dinger equation (NLSE) \cite{C6,C7,C8,C9,C10,C11}, which can solve the puzzle of nature, and associated rogue waves (RWs) are the most
interesting theory in plasma physics, and are employed by a number of authors to investigate the modulational instability (MI)
of the DAWs \cite{Kourakis2003,Moslem2011,Bains2013} and DIAWs \cite{Jukui2003,Pakzad2014} in nonlinear and dispersive plasma medium.
Moslem \textit{et al.} \cite{Moslem2011} considered inertial massive dust grains and inertialess non-extensive electrons and ions
to analyze numerically the condition of the MI of DAWs in presence of non-extensive plasma species, and observed that
the critical wave number $k_c$ increases or decreases according to the sign and magnitude of the $q$. Bains \textit{et al.} \cite{Bains2013} demonstrated
the MI of the DAWs in a plasma having non-extensive electrons and ions. Pakzad \textit{et al.} \cite{Pakzad2014} studied the stability criterion of the
DIAWs in a three component DPM by considering inertial ions, inertialess electrons and immobile massive dust grains.

In DAWs, the moment of inertia is provided by the dust grains and restoring force is provided by the thermal pressure of the
ions and electrons. On the other hand in DIAWs, the moment of inertia is provided by the ions and restoring force
is provided by the thermal pressure of the electrons in presence of immobile dust grains. The mass and charge of the dust grains
are considerably larger than the ions while the mass and charge of the ion are considerably larger
than the electron. It may be noted here that in the DIAWs, if anyone consider
the pressure term of the ions then it is important to be considered the moment of inertia of the ions along with
the dust grains in presence of inertialess electrons. This means that the consideration of the pressure term
of the ions highly contributes to the moment of inertia along with inertial dust grains to generate DIAWs in a DPM
having inertialess electrons. In the present work, we are interested to investigate the nonlinear propagation of DIA RWs (DIARWs) in which the moment
of inertia is provided by the inertial warm ions and negatively charged dust grains and the restoring force is
provided by the thermal pressure of the inertialess electrons in a three component DPM by using the standard NLSE.
\section{Model Equations}
\label{2sec:Model Equations}
We consider a three component DPM comprising of inertial positively charged warm ion (charge $q_i=Z_ie$
and mass $m_i$) and inertial negatively charged dust grains (charge $q_d=-Z_de$ and mass $m_d$) as
well as inertialess non-extensive electron (charge $q_e=-e$; mass $m_e$);
where $Z_i$ ($Z_d$) is the number of  protons (electrons) residing onto the ion (dust grains)
surface, and $e$ is the magnitude of the charge of an electron.
Overall, the charge neutrality condition for our plasma model can
be written as $ Z_in_{i0} = Z_d n_{d0}+ n_{e0}$.  Now, the normalized governing equations
of the DIAWs can be written as
\begin{eqnarray}
&&\hspace*{-1.3cm}\frac{\partial n_d}{\partial t}+ \frac{\partial}{\partial x}(n_d u_d)=0,
\label{2eq:1}\\
&&\hspace*{-1.3cm}\frac{\partial u_d}{\partial t}+u_d \frac{\partial u_d}{\partial x}=\alpha_1 \frac{\partial\phi}{\partial x},
\label{2eq:2}\\
&&\hspace*{-1.3cm}\frac{\partial n_i}{\partial t}+ \frac{\partial}{\partial x}(n_i u_i)=0,
\label{2eq:3}\\
&&\hspace*{-1.3cm}\frac{\partial u_i}{\partial t}+u_i \frac{\partial u_i}{\partial x}+\alpha_2 n_i\frac{\partial n_i}{\partial x}=- \frac{\partial\phi}{\partial x},
\label{2eq:4}\\
&&\hspace*{-1.3cm}\frac{\partial^2\phi}{\partial x^2}+n_i=\alpha_3 n_e+(1-\alpha_3)n_d,
\label{2eq:5}
\end{eqnarray}
where $n_d$ $(n_i)$ is the dust (ion) number density normalized by its equilibrium
value $n_{d0} $ $(n_{i0})$; $u_d$ $(u_i)$ is the dust (ion) fluid speed normalized by
the ion-acoustic wave speed $C_i=(Z_i k_BT_e/m_i)^{1/2}$ with $T_e$ being the non-extensive
electron temperature, and $k_B$ being the Boltzmann constant; $\phi$ is the electrostatic wave potential normalized
by $k_BT_e/e$; the time and space variables are normalized by ${\omega^{-1}_{pi}}=(m_i/4\pi {Z_i}^2 e^2 n_{i0})^{1/2}$
and $\lambda_{Di}=(k_BT_e/4 \pi Z_i e^2 n_{i0})^{1/2}$, respectively. The pressure term of the ion is recognized as
$P_i=P_{i0}(N_i/n_{i0})^\gamma$ with $P_{i0}=n_{i0}k_BT_i$ being the equilibrium
pressure of the ion, and $T_i$ being the temperature of warm ion, and
$\gamma=(N+2)/N$ (where $N$ is the degree of freedom and for one-dimensional case
$N=1$, then $\gamma=3$). Other plasma parameters are
$\alpha_1=\mu\nu$, $\mu=m_i/m_d$, $\nu=Z_d/Z_i$,
$\alpha_2=3T_i/Z_iT_e$, and $\alpha_3=n_{e0}/Z_in_{i0}$, etc.
Now, the expression for the number density of  non-extensive electrons following the
non-extensive distribution \cite{Akter2015,Bacha2012} can be written as
\begin{eqnarray}
&&\hspace*{-1.3cm}n_e=\left[1 + (q-1)\phi \right]^{\frac{(q+1)}{2(q-1)}},
\label{2eq:6}\
\end{eqnarray}
where the parameter $q$ (known as entropic index) quantifies the degree of non-extensivity.
Now, by substituting  Eq. \eqref{2eq:6} into Eq. \eqref{2eq:5}, and expanding up to third order of $\phi$, we get
\begin{eqnarray}
&&\hspace*{-1.3cm}\frac{\partial^2\phi}{\partial x^2}+n_i=\alpha_3+(1-\alpha_3)n_d+M_1\phi+M_2\phi^2+M_3\phi^3+\cdot\cdot\cdot,
\label{2eq:7}
\end{eqnarray}
where
\begin{eqnarray}
&&\hspace*{-1.3cm}M_1 = \alpha_3(1+q)/2,
\nonumber\\
&&\hspace*{-1.3cm}M_2 =\alpha_3(1+q)(3-q)/8,
\nonumber\\
&&\hspace*{-1.3cm}M_3 =\alpha_3(1+q)(q-3)(3q-5)/48.
\nonumber\
\end{eqnarray}
It may be noted here that the terms containing $M_1$, $M_2$, and $M_3$ in the right hand side of
Eq. \eqref{2eq:7} are the contribution of $q$-distributed electrons.
\section{Derivation of NLSE}
\label{2sec:Derivation of NLSE}
To study the MI of the DIAWs, we want to derive the NLSE by employing the reductive perturbation method,
and for that case, first we can write the stretched co-ordinates in the following
form \cite{Kourakis2003,Moslem2011,Bains2013,Jukui2003,Pakzad2014}
\begin{eqnarray}
&&\hspace*{-1.3cm}\xi={\epsilon}(x-v_g t),
\label{2eq:8}\\
&&\hspace*{-1.3cm}\tau={\epsilon}^2t,
\label{2eq:9}
\end{eqnarray}
where $v_g$ is the group speed and $\epsilon$ is a small parameter.
Then we can write the dependent variables \cite{Kourakis2003,Moslem2011,Bains2013,Jukui2003,Pakzad2014} as
\begin{eqnarray}
&&\hspace*{-1.3cm}n_{d}=1 +\sum_{m=1}^{\infty}\epsilon^{m}\sum_{l=-\infty}^{\infty}n_{dl}^{(m)}(\xi,\tau)~\mbox{exp}[i l(kx-\omega t)],
\label{2eq:10}\\
&&\hspace*{-1.3cm}u_{d}=\sum_{m=1}^{\infty}\epsilon^{m}\sum_{l=-\infty}^{\infty}u_{dl}^{(m)}(\xi,\tau)~\mbox{exp}[i l(kx-\omega t)],
\label{2eq:11}\\
&&\hspace*{-1.3cm}n_{i}=1 +\sum_{m=1}^{\infty}\epsilon^{m}\sum_{l=-\infty}^{\infty}n_{il}^{(m)}(\xi,\tau)~\mbox{exp}[i l(kx-\omega t)],
\label{2eq:12}\\
&&\hspace*{-1.3cm}u_{i}=\sum_{m=1}^{\infty}\epsilon^{m}\sum_{l=-\infty}^{\infty}u_{il}^{(m)}(\xi,\tau)~\mbox{exp}[i l(kx-\omega t)],
\label{2eq:13}\\
&&\hspace*{-1.3cm}\phi=\sum_{m=1}^{\infty}\epsilon^{m}\sum_{l=-\infty}^{\infty}\phi_{l}^{(m)}(\xi,\tau)~\mbox{exp}[i l(kx-\omega t)],
\label{2eq:14}
\end{eqnarray}
where $k$ and $\omega$ are real variables representing the carrier wave number and frequency, respectively.
The derivative operators in the above equations are treated as follows:
\begin{eqnarray}
&&\hspace*{-1.3cm}\frac{\partial}{\partial x}\rightarrow\frac{\partial}{\partial x}+\epsilon\frac{\partial}{\partial\xi},
\label{2eq:15}\\
&&\hspace*{-1.3cm}\frac{\partial}{\partial t}\rightarrow\frac{\partial}{\partial t}-\epsilon v_g \frac{\partial}{\partial\xi}+\epsilon^2\frac{\partial}{\partial\tau}.
\label{2eq:16}
\end{eqnarray}
Now, by substituting Eqs. \eqref{2eq:8}-\eqref{2eq:16}  into  Eqs. \eqref{2eq:1}-\eqref{2eq:4}, and \eqref{2eq:7}, and
collecting the terms containing $\epsilon$, the first order ($m=1$ with $l=1$)  reduced equations can be written as
\begin{eqnarray}
&&\hspace*{-1.3cm}n_{d1}^{(1)}=-\frac{k^2\alpha_1}{\omega^2}\phi_1^{(1)},
\label{2eq:17}\\
&&\hspace*{-1.3cm}u_{d1}^{(1)}=-\frac{k \alpha_1}{\omega}\phi_1^{(1)},
\label{2eq:18}\\
&&\hspace*{-1.3cm}n_{i1}^{(1)}=\frac{k^2}{\omega^2-\alpha_2 k^2}\phi_1^{(1)},
\label{2eq:19}\\
&&\hspace*{-1.3cm}u_{i1}^{(1)}=\frac{\omega k }{\omega^2-\alpha_2 k^2}\phi_1^{(1)},
\label{2eq:20}\\
&&\hspace*{-1.3cm}n_{i1}^{(1)}=k^2\phi_1^{(1)}+M_1\phi_1^{(1)}+(1-\alpha_3)n_{d1}^{(1)},
\label{2eq:21}\
\end{eqnarray}
these relations provide the dispersion relation of DIAWs
\begin{eqnarray}
&&\hspace*{-1.3cm}\omega^2=\frac{k^2D_1\pm k^2\sqrt{D_1^2-4D_2D_3}}{2D_2},
\label{2eq:22}\
\end{eqnarray}
where $D_1=1+\alpha_1-\alpha_1\alpha_3+\alpha_2 k^2+\alpha_2 M_1$, $D_2=k^2+M_1$,
and $D_3=\alpha_1\alpha_2-\alpha_1\alpha_2\alpha_3$. In Eq. \eqref{2eq:22},
to get real and positive values of $\omega$, the condition $D_1^2>4D_2D_3$ should be
satisfied. The positive and negative signs in Eq. \eqref{2eq:22} corresponds to the fast ($\omega_f$) and slow ($\omega_s$) DIA modes.
The fast DIA mode corresponds to the case in which both inertial dust and ion components oscillate in phase with the inertialess
electrons. On the other hand, the slow DIA mode corresponds to the case in which only one of the inertial components
oscillates in phase with inertialess electrons, but the other inertial component oscillates in anti-phase with
them \cite{Saberiana2017}. The second-order ($m=2$
with $l=1$) equations are given by
\begin{eqnarray}
&&\hspace*{-1.3cm}n_{d1}^{(2)}=-\frac{k^2\alpha_1}{\omega^2}\phi_1^{(2)}-\frac{2ik\alpha_1(v_gk-\omega)}{\omega^3} \frac{\partial \phi_1^{(1)}}{\partial\xi},
\label{2eq:23}\\
&&\hspace*{-1.3cm}u_{d1}^{(2)}=-\frac{k \alpha_1}{ \omega}\phi_1^{(2)} -\frac{i\alpha_1(v_gk-\omega)}{\omega^2} \frac{\partial \phi_1^{(1)}}{\partial\xi},
\label{2eq:24}\\
&&\hspace*{-1.3cm}n_{i1}^{(2)}=\frac{k^2}{\omega^2-\alpha_2 k^2}\phi_1^{(2)} +\frac{2i\omega k(v_gk-\omega)}{(\omega^2-\alpha_2 k^2)^2} \frac{\partial \phi_1^{(1)}}{\partial\xi},
\label{2eq:25}\\
&&\hspace*{-1.3cm}u_{i1}^{(2)}=\frac{\omega k }{\omega^2-\alpha_2 k^2}\phi_1^{(2)} +\frac{i(v_gk-\omega)(\omega^2+\alpha_2 k^2)}{i(\omega^2-\alpha_2 k^2)^2} \frac{\partial \phi_1^{(1)}}{\partial\xi},
\label{2eq:26}\
\end{eqnarray}
with the compatibility condition
\begin{eqnarray}
&&\hspace*{-1.3cm}v_g=\frac{\alpha_1\alpha_3\omega^5-\alpha_1\omega^5+2\alpha_1\alpha_2\omega^3 k^2-D_4}
{\alpha_1\alpha_3\omega^4k-\alpha_1\omega^4k+2\alpha_1\alpha_2\omega^2 k^3-D_5},
\label{2eq:27}
\end{eqnarray}
where
\begin{eqnarray}
&&\hspace*{-1.3cm} D_4=\alpha_1\alpha_2^2 k^4\omega-\alpha_1\alpha_2^2\alpha_3 k^4\omega-\alpha_2^2\omega^3 k^4
\nonumber\\
&&\hspace*{-0.4cm}-\omega^7+2\alpha_2\omega^5 k^2+\omega^5+2\alpha_1\alpha_2\alpha_3\omega^3 k^2,
\nonumber\\
&&\hspace*{-1.3cm} D_5=2\alpha_1\alpha_2\alpha_3\omega^2 k^3+\alpha_1\alpha_2^2k^5-\alpha_1\alpha_2^2\alpha_3k^5+\omega^4 k.
\nonumber\
\end{eqnarray}
The coefficients of $\epsilon$ for $m=2$ with $l=2$ provide the second
order harmonic amplitudes which are found to be proportional to $|\phi_1^{(1)}|^2$
\begin{eqnarray}
&&\hspace*{-1.3cm}n_{d2}^{(2)}=M_4|\phi_1^{(1)}|^2,
\label{2eq:28}\\
&&\hspace*{-1.3cm}u_{d2}^{(2)}=M_5 |\phi_1^{(1)}|^2,
\label{2eq:29}\\
&&\hspace*{-1.3cm}n_{i2}^{(2)}=M_6|\phi_1^{(1)}|^2,
\label{2eq:30}\\
&&\hspace*{-1.3cm}u_{i2}^{(2)}=M_7 |\phi_1^{(1)}|^2,
\label{2eq:31}\\
&&\hspace*{-1.3cm}\phi_{2}^{(2)}=M_8 |\phi_1^{(1)}|^2,
\label{2eq:32}\
\end{eqnarray}
where
\begin{eqnarray}
&&\hspace*{-1.3cm}M_4=\frac{3k^4\alpha_1^2-2k^2\alpha_1\omega^2M_8 }{2\omega^4},
\nonumber\\
&&\hspace*{-1.3cm}M_5=\frac{k^3\alpha_1^2-2k\omega^2\alpha_1M_8}{2\omega^3},
\nonumber\\
&&\hspace*{-1.3cm}M_6=\frac{2k^2 M_8(\omega^2-\alpha_2 k^2)^2+k^4(\alpha_2 k^2+3\omega^2)}{2(\omega^2-\alpha_2 k^2)^3},
\nonumber\\
&&\hspace*{-1.3cm}M_7=\frac{ \omega^3M_6-\alpha_2\omega k^2M_6-\omega k^4}{k(\omega^2-\alpha_2 k^2)^2},
\nonumber\\
&&\hspace*{-1.3cm}M_8=\frac{D_6-2\omega^4 M_2(\omega^2-\alpha_2 k^2)^3}{6k^2\omega^4(\omega^2-\alpha_2 k^2)^3},
\nonumber\
\end{eqnarray}
where $D_6=k^4\omega^4(3\omega^2+\alpha_2 k^2)-3\alpha_1^2 k^4(\omega^2-\alpha_2 k^2)^3$.
Now, we consider the expression for ($m=3$ with $l=0$) and ($m=2$ with $l=0$),
which leads the zeroth harmonic modes. Thus, we obtain
\begin{eqnarray}
&&\hspace*{-1.3cm}n_{d0}^{(2)}=M_9|\phi_1^{(1)}|^2,
\label{2eq:33}\\
&&\hspace*{-1.3cm}u_{d0}^{(2)}=M_{10}|\phi_1^{(1)}|^2,
\label{2eq:34}\\
&&\hspace*{-1.3cm}n_{i0}^{(2)}=M_{11}|\phi_1^{(1)}|^2,
\label{2eq:35}\\
&&\hspace*{-1.3cm}u_{i0}^{(2)}=M_{12}|\phi_1^{(1)}|^2,
\label{2eq:36}\\
&&\hspace*{-1.3cm}\phi_0^{(2)}=M_{13} |\phi_1^{(1)}|^2,
\label{2eq:37}\
\end{eqnarray}
where
\begin{eqnarray}
&&\hspace*{-1.3cm}M_{9}=\frac{2v_g\alpha_1^2 k^3+\alpha_1^2k^2\omega-\alpha_1\omega^3M_{13}}{v_g^2\omega^3},
\nonumber\\
&&\hspace*{-1.3cm}M_{10}=\frac{\alpha_1^2k^2-\alpha_1\omega^2M_{13}}{v_g\omega^2},
\nonumber\\
&&\hspace*{-1.3cm}M_{11}=-\frac{M_{13}(\omega^2-\alpha_2k^2)^2+k^2(\alpha_2k^2+\omega^2+2\omega v_g k)}{(\alpha_2-v_g^2)(\omega^2-\alpha_2k^2)^2},
\nonumber\\
&&\hspace*{-1.3cm}M_{12}=\frac{v_gM_{11}(\omega^2-\alpha_2k^2)^2-2\omega k^3}{(\omega^2-\alpha_2k^2)^2},
\nonumber\\
&&\hspace*{-1.3cm}M_{13}= \frac{2M_2v_g^2\omega^3(\alpha_2-v_g^2)(\omega^2-\alpha_2k^2)^2+D_7}{\omega^3\times D_8\times(\omega^2-\alpha_2k^2)^2},
\nonumber\
\end{eqnarray}
where
\begin{eqnarray}
&&\hspace*{-1.3cm}D_7=(1-\alpha_3)(\alpha_2-v_g^2)(\omega^2-\alpha_2k^2)(\alpha_1^2k^2\omega+2v_g\alpha_1^2 k^3)
\nonumber\\
&&\hspace*{-0.4cm}+k^2v_g^2\omega^3(\alpha_2k^2+\omega^2+2\omega v_g k),
\nonumber\\
&&\hspace*{-1.3cm}D_8=[\alpha_1(1-\alpha_3)(\alpha_2-v_g^2)-v_g^2-v_g^2M_1(\alpha_2-v_g^2)].
\nonumber\
\end{eqnarray}
Finally, the third harmonic modes ($m=3$) and ($l=1$) with the help of Eqs. \eqref{2eq:17}$-$\eqref{2eq:37},
give a set of equations, which can be reduced to the following NLSE:
\begin{eqnarray}
&&\hspace*{-1.3cm}i\frac{\partial\Phi}{\partial\tau}+P\frac{\partial^2\Phi}{\partial\xi^2}+Q|\Phi|^2\Phi=0,
\label{2eq:38}
\end{eqnarray}
where $\Phi=\phi_1^{(1)}$ for simplicity. In Eq. \eqref{2eq:38}, $P$ is the dispersive coefficient which can be written as
\begin{eqnarray}
&&\hspace*{-1.3cm}P=\frac{\omega^4(\omega-v_g k)^2(\omega^2+\alpha_2 k^2)-D_9}{\omega (\omega^2-\alpha_2 k^2)[2\omega^4 k^2+2\alpha_1 k^2(1-\alpha_3)(\omega^2-\alpha_2 k^2)^2]},
\nonumber\
\end{eqnarray}
where
\begin{eqnarray}
&&\hspace*{-1.3cm}D_9=2\omega^5 kv_g(\omega-v_gk)(\omega v_g-\alpha_2 k)+\omega^4(\omega^2-\alpha_2 k^2)^3
\nonumber\\
&&\hspace*{-0.5cm}+\alpha_1(1-\alpha_3)(\omega-v_g k)(3v_gk-\omega)(\omega^2-\alpha_2 k^2)^3,
\nonumber\
\end{eqnarray}
and also $Q$ is the nonlinear coefficient which can be written as
\begin{eqnarray}
&&\hspace*{-1.3cm}Q=\frac{\omega^3(\omega^2-\alpha_2 k^2)^2[3M_3+2M_2(M_8+M_{13})]-D_{10}}{2\omega^4 k^2+2\alpha_1 k^2(1-\alpha_3)(\omega^2-\alpha_2 k^2)^2},
\nonumber\
\end{eqnarray}
where
\begin{eqnarray}
&&\hspace*{-1.3cm}D_{10}=\alpha_1\omega k^2(1-\alpha_3)(\omega^2-\alpha_2 k^2)^2(M_4+M_9)
\nonumber\\
&&\hspace*{-0.4cm}+2\alpha_1k^3(1-\alpha_3)(\omega^2-\alpha_2 k^2)^2(M_5+M_{10})
\nonumber\\
&&\hspace*{-0.4cm}+\omega^3(\omega^2k^2+\alpha_2k^4)(M_6+M_{11})+2\omega^4k^3                                                                     (M_7+M_{12}).
\nonumber\
\end{eqnarray}
The space and time evolution of the DIAWs in DPM are directly governed by the coefficients $P$ and $Q$,
and indirectly governed by different plasma parameters such as $\mu$, $\nu$, $\alpha_2$, $\alpha_3$, $q$, and $k$.
Thus, these plasma parameters significantly
affect the stability conditions of the DIAWs.
\section{Modulational instability and Rogue waves}
\label{2sec:Modulational instability and Rogue waves}
The stable and unstable parametric regimes of the DIAWs are organized by the sign of $P$ and $Q$ of the
standard NLSE  \eqref{2eq:38} \cite{Kourakis2003,Moslem2011,Bains2013,Jukui2003,Pakzad2014,Li2011,Li2017,Liu2016,Liu2017}.
When $P$ and $Q$ have same sign (i.e., $P/Q>0$), the evolution of the DIAWs amplitude is
modulationally unstable in presence of the external perturbations. On the other hand,
when $P$ and $Q$ have opposite sign (i.e., $P/Q<0$), the DIAWs are modulationally stable in presence of the external perturbations.
The plot of $P/Q$ against $k$ yields stable and unstable parametric regimes of the DIAWs.
The point, at which transition of $P/Q$ curve intersect with $k$-axis, is known as threshold
or critical wave number $k$ ($=k_c$). The governing equation of the highly energetic DIARWs
in the unstable parametric regime (i.e., $P/Q>0$) can be written as \cite{Ankiewiez2009}
\begin{eqnarray}
&&\hspace*{-1.0cm}\Phi(\xi,\tau)=\sqrt{\frac{2P}{Q}}\Big[\frac{4+16 i\tau P}{1+4 \xi^2 + 16\tau^2 P^2}-1\Big] \mbox{exp}(2i\tau P),
\label{2eq:39}
\end{eqnarray}
\begin{figure}[t!]
\centering
\includegraphics[width=80mm]{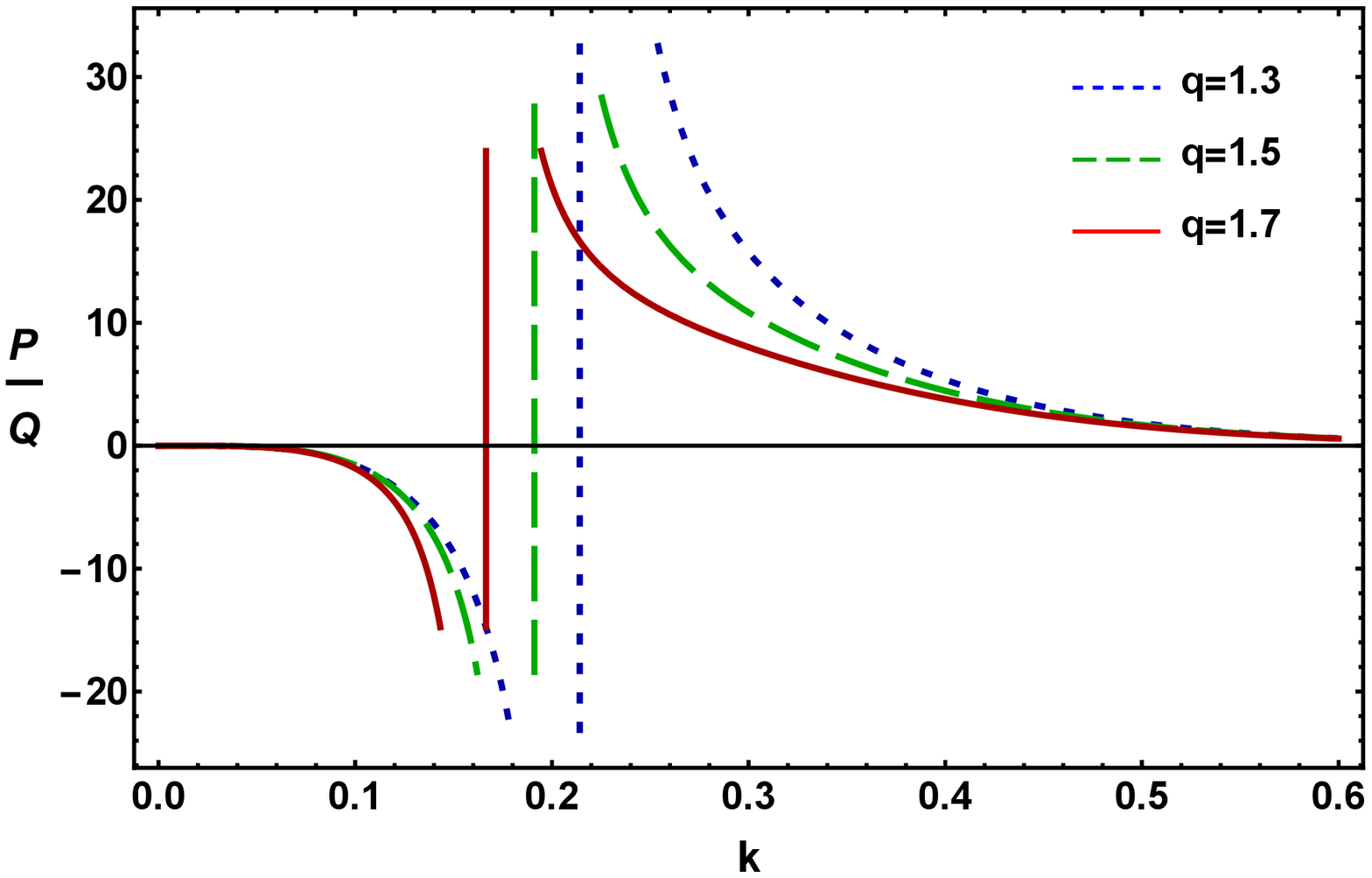}
\caption{Plot of $P/Q$ vs $k$ for different values of positive $q$ when other plasma parameters are $\mu=2\times 10^{-6}$, $\nu=3\times 10^{5}$, $\alpha_2=0.5$, $\alpha_3=0.2$, and $\omega_f$.}
\label{2Fig:F1}
\vspace{0.8cm}
\includegraphics[width=80mm]{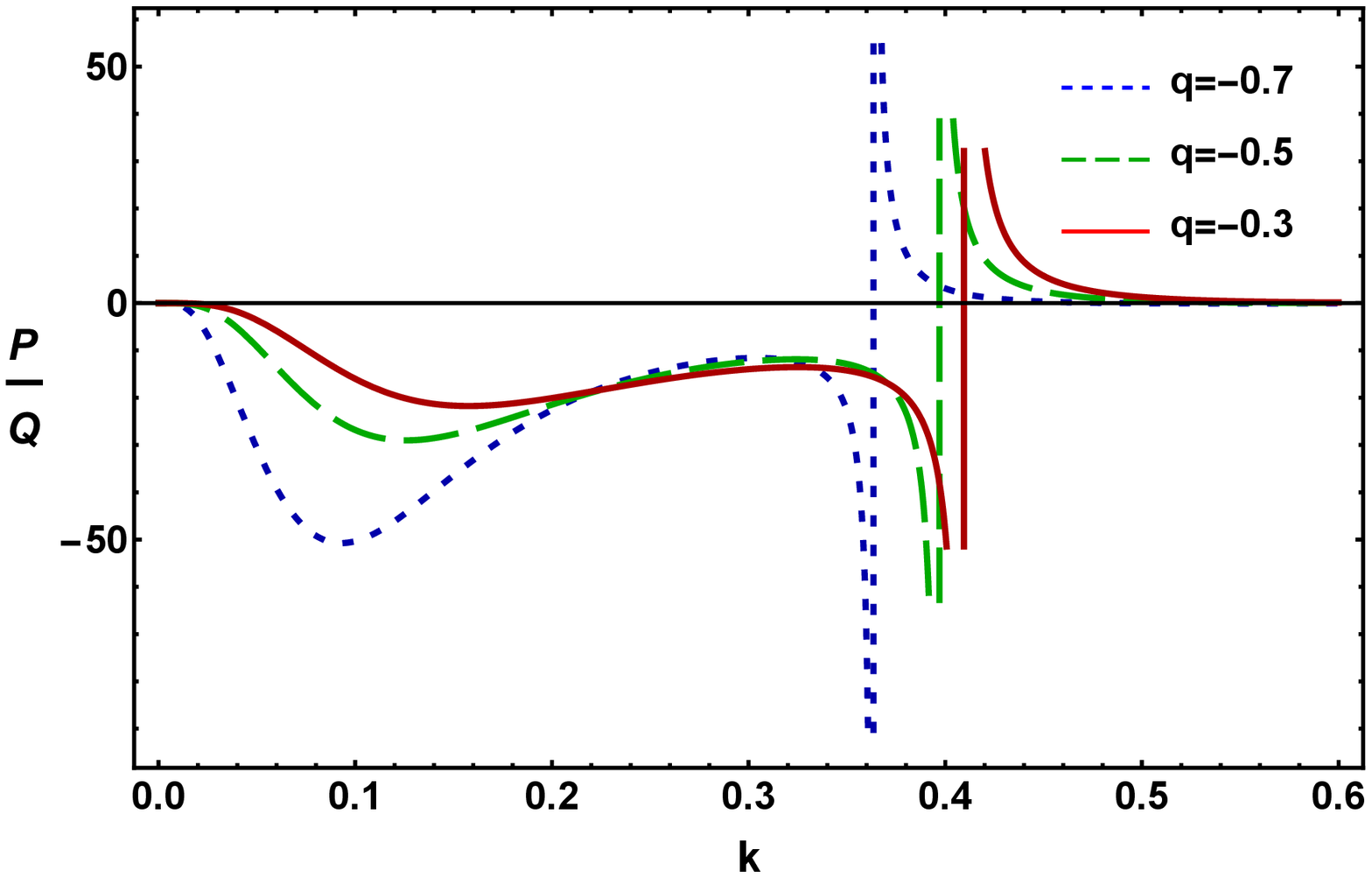}
\caption{Plot of $P/Q$ vs $k$ for different values of negative $q$ when other plasma parameters are $\mu=2\times 10^{-6}$, $\nu=3\times 10^{5}$, $\alpha_2=0.5$, $\alpha_3=0.2$, and $\omega_f$.}
 \label{2Fig:F2}
\end{figure}
\begin{figure}[t!]
\centering
\includegraphics[width=80mm]{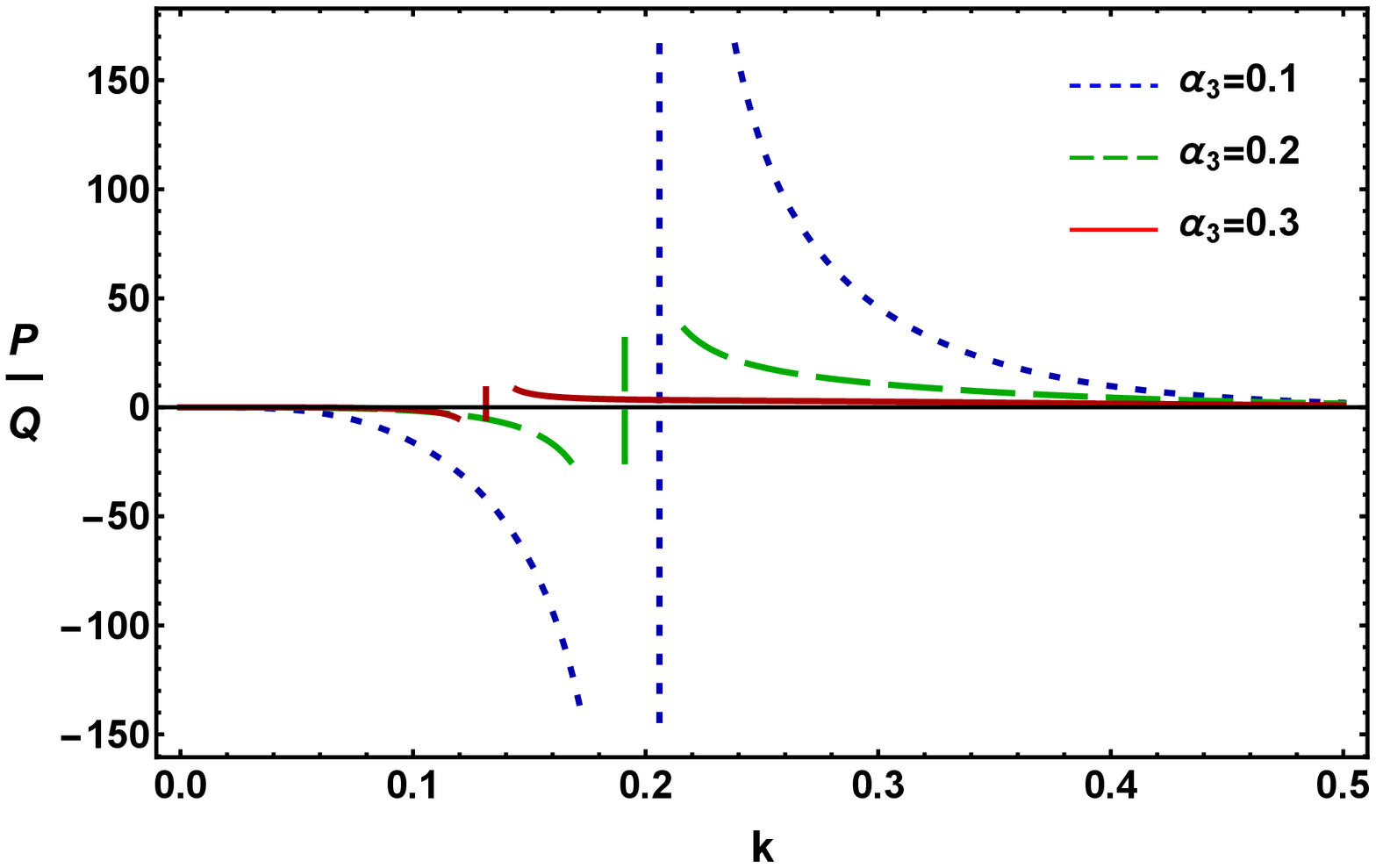}
\caption{Plot of $P/Q$ vs $k$ for different values of positive $\alpha_3$ when other plasma parameters are $q=1.5$ $\mu=2\times 10^{-6}$, $\nu=3\times 10^{5}$, $\alpha_2=0.5$, and $\omega_f$.}
\label{2Fig:F3}
\end{figure}
\begin{figure}[t!]
\centering
\includegraphics[width=80mm]{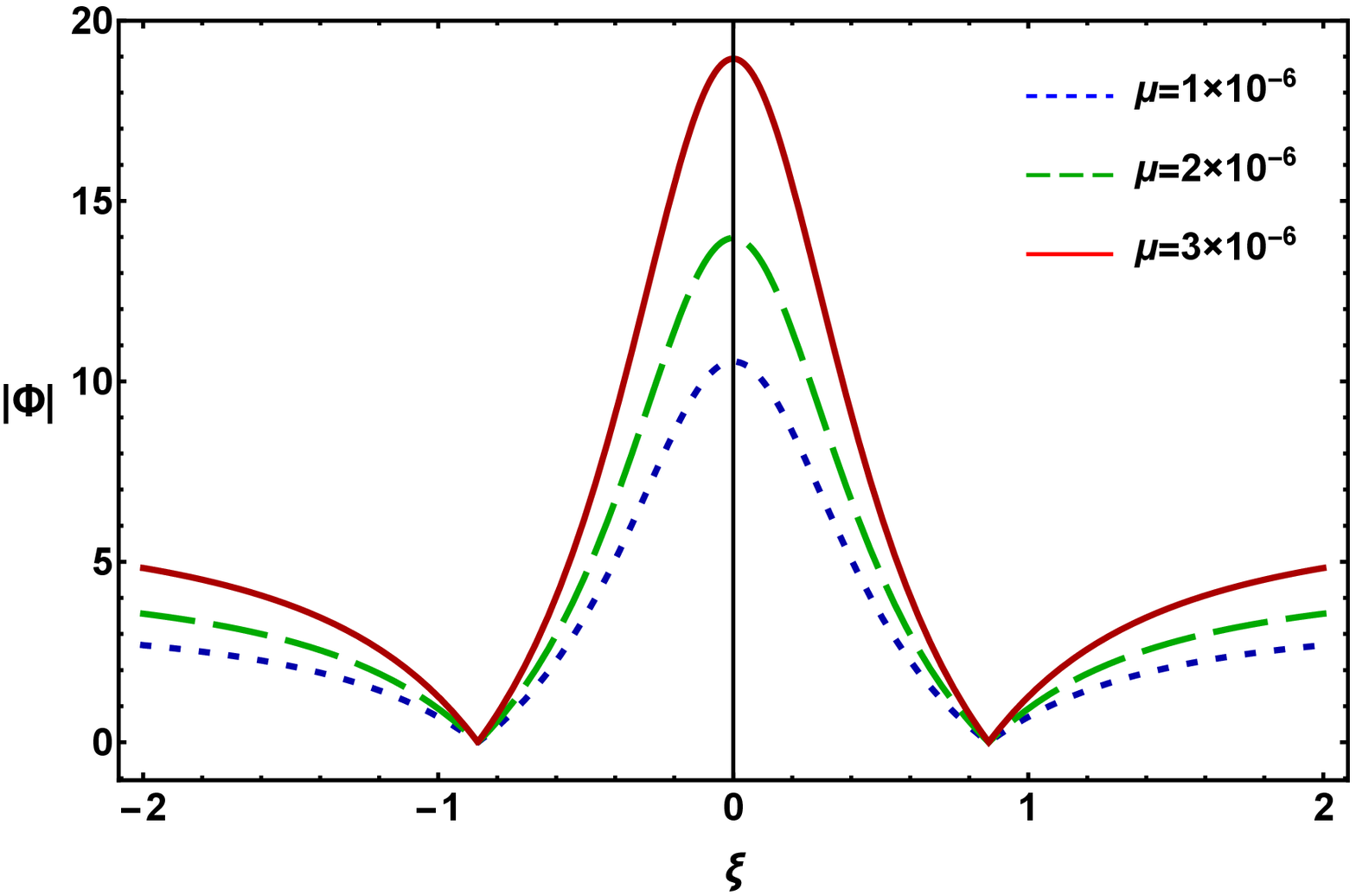}
\caption{Plot of $Re(\Phi)$ vs $\xi$ for different values of $\mu$ when
$k=0.3$, $\tau=0.0$, $q=1.5$, $\nu=3\times 10^{5}$, $\alpha_2=0.5$, $\alpha_3=0.2$, and $\omega_f$.}
\label{2Fig:F4}
\end{figure}
\begin{figure}[t!]
\centering
\includegraphics[width=80mm]{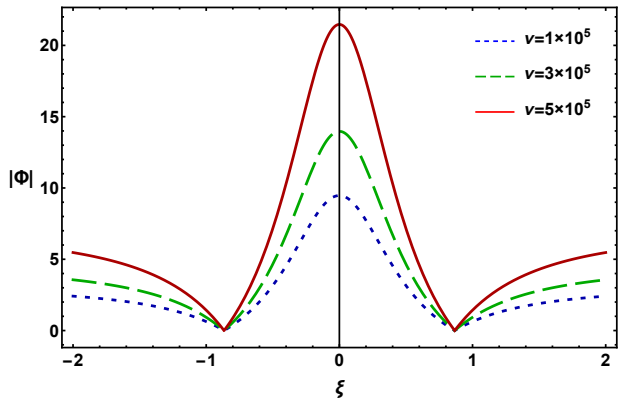}
\caption{Plot of $Re(\Phi)$ vs $\xi$ for different values of $\nu$ when $k=0.3$, $\tau=0.0$,
$q=1.5$,  $\mu=2\times 10^{-6}$, $\alpha_2=0.5$, $\alpha_3=0.2$, and $\omega_f$.}
\label{2Fig:F5}
\end{figure}
Equation \eqref{2eq:39} describes that a large amount of wave energy, which causes due to the
nonlinear characteristics of the medium, is localized into a comparatively small area in space.
\section{Results and discussion}
\label{2sec:Results and discussion}
Now, we would like to numerically analyze the stability conditions of
the DIAWs in presence of the non-extensive  electrons. The mass and charge state of the
plasma species even their number density are important factors in recognizing the stability
conditions of the DIAWs in DPM \cite{Saberiana2017,Merlino2014,Shukla2012,Mamun2002,Shalaby2009,Shukla2002}.
The mass of the dust grains is comparable to the mass of proton. In general picture of the
DPM, dust grains are massive (millions to billions times heavier than the protons) and their
sizes range from nanometres to millimetres. Dust grains may be metallic, conducting, or made of ice
particulate. The size and shape of dust grains will be different, unless they are man-made.
The dust grains are millions to billions times heavier than the protons, and typically, a dust grain acquires one
thousand to several hundred thousand elementary charges \cite{Saberiana2017,Merlino2014,Shukla2012,Mamun2002,Shalaby2009,Shukla2002}.
In this article, we consider three components dusty plasma model having inertial warm positive ions and negative dust grains,
and interialess non-thermal non-extensive electrons. It may be noted here that in the DIAWs, if anyone consider
the pressure term of the ions then it is important to be considered the moment of inertia of the ions along with
the dust grains in presence of inertialess electrons. This means that the consideration of the pressure term
of the ions highly contributes to the moment of inertia along with inertial dust grains
to generate DIAWs in a DPM having inertialess electrons. In our present analysis, we have
considered that $m_d=10^6m_i$, $Z_d=(10^3\thicksim10^5)Z_i$, and $T_e=10T_i$.

The detail picture of the non-extensive parameter (for both sub-extensive and super-extensive
limits) in recognizing stable and unstable parametric regimes of the DIAWs can be observed in
Figs. \ref{2Fig:F1} and \ref{2Fig:F2} for $\omega_f$ only, and it is obvious from this two
figures that (a) both modulationally stable (i.e., $P/Q<0$) and unstable (i.e., $P/Q>0$) parametric
regimes of the DIAWs can be obtained for sub-extensive (i.e., $q>1$) and super-extensive (i.e., $q<1$)
limits of $q$; (b) the $k_c$ decreases with the increase in the value of $q$ within the limit of
sub-extensive (i.e., $q>1$) but the $k_c$ increases with the increase in the value of $q$ within the
limit of super-extensive (i.e., $q<1$); (c) so, the direction of the variation of $k_c$ totally
depends on the sign of the $q$.

The influence of the charge state of positive ion, and the number density
of the warm ions and electrons on the possible unstable and stable parametric regimes of DIAWs can be seen
from Fig. \ref{2Fig:F3} in which the variation of the $P/Q$ with $k$ for different values of
$\alpha_3$ is shown when other plasma parameters are remain constant. It is clear from this figure
that (a) when the parameter $\alpha_3$ is $0.1$, $0.2$, and $0.3$, then the corresponding
$k_c$ is almost $0.21$, $0.20$, and $0.13$; (b) so, the stable window  as well as the $k_c$
decreases with increasing $\alpha_3$; (c) the presence of excess number of non-extensive
electrons dictates the DIAWs to be unstable as well allows to generate DIARWs for small
wave number while the presence of excess number of warm ions dictates the DIAWs to be unstable as well allows
to generate DIARWs for large wave number for a constant value of ion charge state (via $\alpha_3$).

The effects of inertial warm positive ion mass to inertial negatively charged massive dust grains mass
(via $\mu$) on the formation of the DIARWs can be observed from Fig. \ref{2Fig:F4}, and it can be highlighted
from this figure that (a) the increase in the value of $\mu$ does not only cause to change the nonlinearity of the
plasma medium but also causes to increase the amplitude and width of the DIARWs; (b) physically, the nonlinearity
of the plasma medium as well as the amplitude and width of the DIARWs increases (decreases) with the increase of positive
ion mass (negative dust mass). Figure \ref{2Fig:F5} demonstrates the influence of the ratio of the charge
state of negative dust grains to the charge state of positive ion in recognizing the shape of the DIARWs in a nonlinear and
dispersive plasma medium (via $\nu$). The amplitude and width of the DIARWs associated with modulationally unstable parametric regime
decreases (due to the decrease of the nonlinearity of plasma medium) with positive ion charge state while increases (due to the increase
of the nonlinearity of plasma medium) with negative dust grains charge state (via $\nu$).
\section{Conclusion}
\label{2sec:Conclusion}
In this paper, we have considered a dusty plasma model having inertial warm positive ions and
negative dust grains, and inertialess non-extensive electrons to investigate the stability criteria,
which determines by the sign of $P$ and $Q$ of NLSE, of the DIAWs according to the
variation of plasma parameters, and have also observed the formation of DIARWs in the modulationally
unstable parametric regime. The variation of the mass and charge state of the positive ion and
negative dust grains has dictated a rigourous change in the configuration of DIARWs associated with DIAWs in the modulationally
unstable parametric regime. The results from this brief study are to be applicable in explaining the formation of nonlinear
electrostatic DIARWs in space (viz., Earth ionosphere \cite{El-Labany2018}, asteroid zones \cite{Eslami2013}, magnetosphere \cite{El-Labany2018},
protostellar disks \cite{Eslami2013}, molecular clouds \cite{Eslami2013}, interstellar medium \cite{El-Labany2018},
cometary tails \cite{Eslami2013}, and planetary rings \cite{El-Labany2018},  etc.) and laboratory plasmas.

\end{document}